\documentclass{ar-1col}
\usepackage[numbers]{natbib}
\usepackage{url}
\usepackage{amsmath,amssymb}
\usepackage{multicol}
\definecolor{green}{rgb}{0,0.5,0}
\usepackage{tabularx}
\usepackage{mathtools}
\usepackage{bm}
\usepackage{soul}
\usepackage{siunitx}
\usepackage{comment}
\usepackage[normalem]{ulem}
\usepackage[super]{nth}
\usepackage[T1]{fontenc}

\jname{Xxxx. Xxx. Xxx. Xxx.}
\jvol{AA}
\jyear{YYYY}
\doi{10.1146/((please add article doi))}

\newcommand{\figref}[1]{\figurename\,\ref{#1}}

\newcommand{\figpanel}[1]{\textbf{\textsf{#1}}}
\newcommand{\figrefp}[2]{\figurename\,\ref{#1}\,\textbf{\textsf{#2}}}

\newcommand{\secref}[1]{Sec.\,\ref{#1}}
\newcommand{\Secref}[1]{Section\,\ref{#1}}

\newcommand{\phij}{\phi_\mathrm{J}}
\newcommand{\phig}{\phi_\mathrm{G}}
\newcommand{\nano}{\si{\nano\metre}}
\newcommand{\suspvisc}{\eta}
\newcommand{\fluidvisc}{\eta_\mathrm{f}}
\newcommand{\fluiddens}{\rho_\mathrm{f}}
\newcommand{\partdens}{\rho_\mathrm{p}}

\newcommand{\gd}{\dot{\gamma}}
\newcommand{\phim}{\phi_{\mathrm{m}}}
\newcommand{\ReNum}{\mbox{\textit{Re}}}
\newcommand{\StNum}{\mbox{\textit{St}}}
\newcommand{\PeNum}{\mbox{\textit{Pe}}}

\DeclareMathAlphabet{\mathsfit}{\encodingdefault}{\sfdefault}{m}{sl}
\SetMathAlphabet{\mathsfit}{bold}{\encodingdefault}{\sfdefault}{bx}{sl}
\newcommand{\tensor}[1]{\bm{\mathsfit{#1}}}

\begin{document}

\markboth{Ness {\arblcirc} Seto {\arblcirc} Mari}{Physics of dense suspensions}
\title{The physics of dense suspensions}

\author{Christopher Ness,$^1$ Ryohei Seto,$^{2,3}$ and Romain Mari$^4$
\affil{$^1$School of Engineering, University of Edinburgh, Edinburgh EH9 3FG, United Kingdom; email: chris.ness@ed.ac.uk}
\affil{$^2$Wenzhou Institute, University of Chinese Academy of Sciences, Wenzhou, Zhejiang 325000, China; 
email: seto@wiucas.ac.cn}
\affil{$^3$Oujiang Laboratory, Wenzhou, Zhejiang 325000, China}
\affil{$^4$Universit\'e Grenoble Alpes, CNRS, Laboratoire Interdisciplinaire de Physique (LIPhy), 38000 Grenoble, France; email: romain.mari@univ-grenoble-alpes.fr}}

\begin{abstract}
Dense suspensions of particles are relevant to many applications and are a key platform for developing a fundamental physics of out-of-equilibrium systems.
They present challenging flow properties, apparently turning from liquid to solid upon small changes in composition or, intriguingly, in the driving forces applied to them. 
The emergent physics close to the ubiquitous jamming transition (and to some extent the glass and gelation transitions) provides common principles with which to achieve a consistent interpretation of a vast set of phenomena reported in the literature. 
In light of this, we review the current state of understanding regarding the relation between the physics at the particle scale and the rheology at the macroscopic scale.
We further show how this perspective opens new avenues for the development of continuum models for dense suspensions.
\end{abstract}

\maketitle

\section{INTRODUCTION}
\label{sec:intro}
\begin{figure}
\includegraphics[trim = 0mm 175mm 0mm 0mm, clip,width=1\textwidth,page=1]{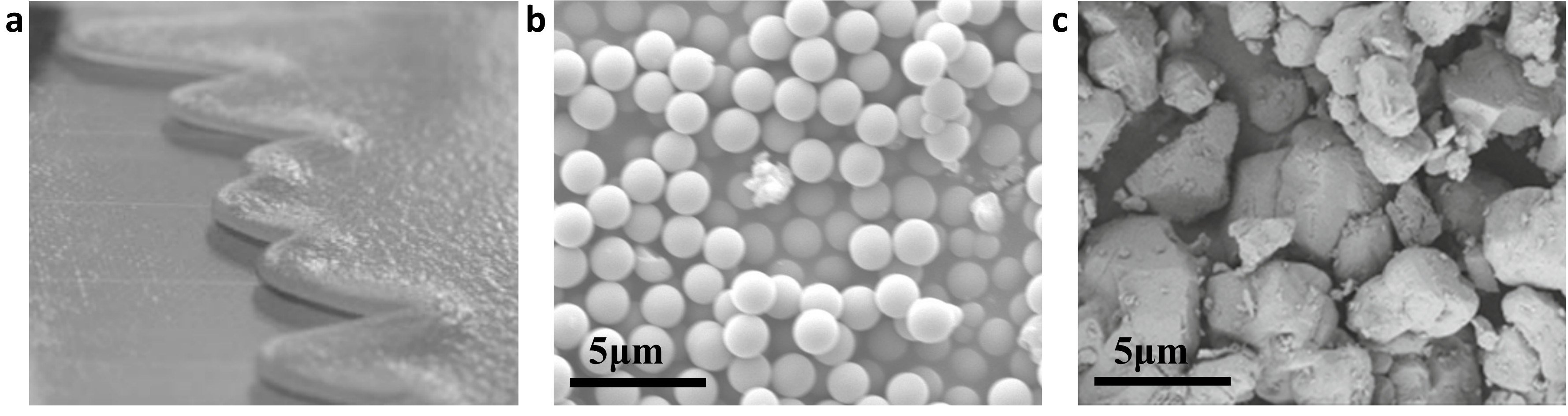}
\caption{
Dense suspensions.
\figpanel{a} A particle-laden film comprising \SIrange{250}{450}{\micro\metre} glass beads in glycerol~\cite{zhou2005theory}.
Figure reprinted with permission from Zhou et al.,
Physical Review Letters 94:117803 (2005);
copyright (2005) the American Physical Society;
\figpanel{b} a model system of \SI{1.4}{\micro\metre} monodisperse silica spheres~\cite{fusier2018rheology}.
Figure reprinted with permission from Fusier et al.,
Journal of Rheology 62(3):753--771 (2018);
copyright (2018) The Society of Rheology;
\figpanel{c} calcium carbonate particles of approximately $\SI{5}{\micro\metre}$, with irregular shape~\cite{bossis2017discontinuous}.
Figure adapted with permission from Bossis et al.,
Rheologica Acta 56:415--430 (2017);
copyright (2017) Rheologica Acta;
}
\label{figure1}
\end{figure}

Suspensions of particles in liquid are found throughout nature and industry,
with examples ranging from mud, magma and blood to cement, paint and molten chocolate.
Often solid and fluid are mixed in roughly equal proportion,
leading to a thick or ``pasty'' consistency, see~\figref{figure1}\,\figpanel{a}.
The widespread use of these dense suspensions is enabled by
extensive experimental characterization and empirical modelling
of their mechanical behavior, 
or \emph{rheology}, 
allowing one to estimate e.g.\ the thickness of a poured coating, the energy required to stir a slurry, or the extent of a mudslide.
Nonetheless, there are so far no accepted continuum theories; we have not yet identified the dense suspension analog of the Navier--Stokes equations.
Doing so requires us to understand the relationship between composition and material properties, that is, to develop a physics of suspensions.

This effort has a rich and extensive history,
dating back at least to Einstein~\cite{einstein1911berichtigung}. 
The physics of suspensions has long been seen primarily as a fluid mechanical problem, in which the dynamics are dominated by viscous stresses induced by the presence of particles.
This perspective led to Batchelor's successful theory for dilute suspensions of solid volume fraction $\phi\lesssim 0.1$~\cite{batchelor_stress_1970}.
More concentrated suspensions, however, display behaviors that have proven elusive to the fluid mechanical approach, 
in particular their tendency to transition reversibly from liquid to solid depending on the applied stress. 
Ketchup, for instance, will only flow out of the bottle under \emph{large} stresses,
whereas cornstarch suspensions (c.f.\ Dr.\ Seuss's Oobleck) will only flow smoothly under \emph{small} stresses.

The reminiscence of phase transitions motivated the development of near-equilibrium statistical mechanics approaches (notably mode-coupling theory) to some of these problems in the 2000s, 
often attributing the emergence of solidity to a glass transition (see for instance~\cite{fuchsTheoryNonlinearRheology2002}).
While this approach proved a powerful tool for \emph{colloidal} suspensions, 
in many cases the particles are too large to be significantly influenced by Brownian motion; 
these suspensions are far from equilibrium.
More recently, researchers started to explore an analogy between dense suspensions and dry granular materials, 
which at the macroscopic scale led to a highly influential constitutive law, the ``$\mu(J)$'' rheology~\cite{boyer2011unifying}. 
This further triggered intense interest in the microscopic parallels between these two materials, in particular in the role of particle contact forces and associated friction 
in the stress of dense \emph{granular} suspensions.
The present review centers mostly on the recent developments of this approach, and how a coherent physics of dense suspensions is emerging from it.
As we will see, the rheology of a dense suspension is intimately linked to an underlying out-of-equilibrium transition known as jamming, 
occurring when the solid volume fraction $\phi$ reaches a critical value $\phij$.
By accounting for the proximity of $\phi$ to this critical value,
which, crucially, is stress-dependent through  
many aspects of particle-level physics,
one can understand broad spectra of rheology
in a consistent way.

Dense suspensions are a mainstay of soft matter science:
understanding their physics is crucial in many industrial settings
and is of fundamental relevance.
This (necessarily short and incomplete) review aims to equip the reader with an intuitive starting point from which to tackle the vast and varied physics of dense suspensions.
Each of the topics covered deserves its own detailed review\,---\,indeed many of these have been written and will be cited (see e.g.~\cite{mewis2012colloidal,guazzelli2018rheology,morris2020shear,Denn_2014}).
In what follows,
we first outline the pertinent features of dense suspensions.
We then introduce jamming in \Secref{sec:noneq},
before discussing in \Secref{sec:rheology} how the macroscopic rheology is 
governed by microscopic physics that set the proximity to the transition.
In \Secref{sec:const} we show how microscopic physics can be encoded in continuum models,
before introducing some aspects of
the fluid mechanics of dense suspensions.

\subsection{What are dense suspensions?}
\label{sec:susps}

Dense suspensions represent a large subset of complex fluids and can vary considerably in their physical and chemical composition. Here we briefly list the main sources of physical variability and the characteristics that make a comprehensive description of their physics challenging.

\begin{figure}
\includegraphics[trim = 0mm 137mm 0mm 0mm, clip,width=1\textwidth,page=2]{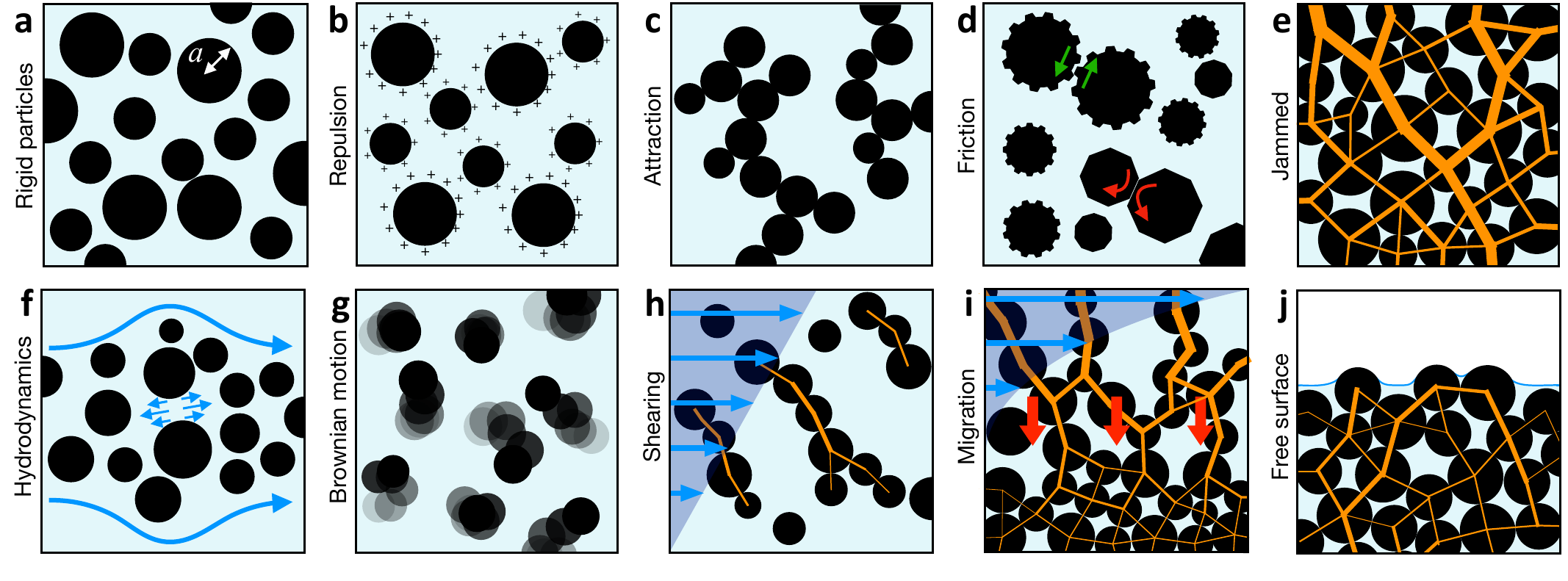}
\caption{
Schematics of dense suspension microphysics.
\figpanel{a} A suspension of rigid particles;
\figpanel{b} repulsive interactions;
\figpanel{c} attractive interactions;
\figpanel{d} particle-particle friction, showing inhibited sliding (rough particles highlighting surface asperities, green arrows) and inhibited rolling (faceted particles, red arrows);
\figpanel{e} a jammed state (force chains in orange);
\figpanel{f} hydrodynamics, showing lubrication (short arrows) and drag (long arrows); 
\figpanel{g} Brownian motion;
\figpanel{h} shear induced structure showing particle contact forces (orange lines) under a shear flow (streaming velocity shown in blue arrows);
\figpanel{i} particle migration (red arrows), under a shear rate gradient (blue arrows) with a gradient in the magnitude of particle contact forces (orange lines);
\figpanel{j} free surface, showing curvature of the interface on the scale of the particle size $a$, leading to capillary forces.
}
\label{figure2}
\end{figure}

\smallskip \noindent
\textbf{\emph{The particles}}
may be stiff solids 
of arbitrary shape (see~\figref{figure1}\,\figpanel{b},\,\figpanel{c})
with crystalline
or amorphous
structure,
or may instead be soft materials such as hydrogel particles
or red blood cells. 
We focus here on particles that can, to a first approximation, be considered rigid,~\figref{figure2}\,\figpanel{a}.
While there are no strict bounds on the particle size $a$,
we restrict the discussion to $\mathcal{O}(100)\si{\nano\metre}$ $\lesssim a\lesssim\mathcal{O}(1)\si{\milli\metre}$.
Such particles are orders of magnitude larger than their molecular constituents and
can, in principle, be characterized using continuum mechanics concepts such as their Young modulus.
Crucially, though, they are usually not \emph{macroscopic} bodies,
and nanometre scale physics remain relevant.
The marriage of macroscopic and microscopic physics distinguishes their interactions from those between molecules.

Surface effects in particular
play a major role in the interactions between particles~\cite{israelachvili2011intermolecular}.
Often surfaces are ``dressed'', either by adsorbed molecules such as polymer brushes   
or by ion/counterion double layers, providing repulsion (\figref{figure2}\,\figpanel{b})
at length scales typically in the $\si{\nano\metre}$ to $\si{\micro\metre}$ range~\cite{mewis2012colloidal}. 
At this scale surface roughness is also present, as are Van der Waals forces~\cite{mewis2012colloidal}.
The former may allow the lubrication singularity to be violated;
the latter may introduce particle attraction (\figref{figure2}\,\figpanel{c}). 
To complicate matters further, concepts from the physics of dry granular contacts, 
for instance friction~(\figref{figure2}\,\figpanel{d}), 
adhesion
and force chains~\cite{Cates_1998a} (\figref{figure2}\,\figpanel{e}),
appear relevant~\cite{seto2013discontinuous}.
Sliding friction in particular has been observed at particle level in systems varying considerably in their composition~\cite{fernandez_microscopic_2013,comtet_pairwise_2017},
though its origin may differ from that between macroscopic bodies.
In both cases, surface morphology probably plays a central role, but whereas for large bodies 
this creates multi-contact interfaces with statistics leading to
Amontons--Coulomb law~\cite{johnson_contact_1985},
for small particles contact through one or few asperities 
may be the norm (dependent on the extent of contact deformation), giving rise 
to load-weakening friction~\cite{chatte2018shear,lobry2019shear}.
%
%
Combining all of these effects,
the situation is sufficiently complex that there are no established force models applicable across the size range of interest:
understanding how particles interact is a challenge.

\smallskip \noindent
\textbf{\emph{The fluid,}}
or solvent, is the main factor differentiating suspensions from dry granular media. 
It mediates hydrodynamic interactions,~\figref{figure2}\,\figpanel{f}, 
and can be a simple liquid (e.g.\ water), 
a polymeric fluid (e.g.\ filled thermoplastics), or a complex, multiphase fluid (e.g.\ the cement paste in fresh concrete).
In general these may have dissolved ions and molecules  that influence the interactions described above, via e.g.\ a change of pH or dielectric constant.
Here we focus on Newtonian suspending fluids
with constant viscosity $\fluidvisc$ and density $\fluiddens$.
The fluid contributes a dissipative stress scale $\fluidvisc \gd$ (for shear rate $\gd$)
and ensures incompressibility of the system, 
which plays a major role at macroscopic scales (e.g.~\cite{Nagahiro_2016}).
Additionally, thermal fluctuations in the fluid velocity generate Brownian forces on particles, which is especially relevant at the lower end of particle size range considered here,~\figref{figure2}\,\figpanel{g}.

\smallskip \noindent
\textbf{\emph{Crowded conditions.}}
Although it has no formal definition, a `dense' suspension is usually taken to be one with
solid and liquid mixed in roughly equal proportion.
Under these conditions, the typical particle separation is smaller than the particle size
and small strains may bring their surfaces into contact (or near-contact).
Short-range interactions such as pairwise repulsion/attraction are therefore important.
Meanwhile long-range, many-body hydrodynamics are screened by intervening particles,
and the emerging consensus is that these are negligible in dense systems.
Lubrication forces also remain relevant at a quantitative level only~\cite{maiti_rheology_2014}.
Increasing $\phi$,
suspensions generically undergo a \emph{jamming} transition, described later, from a flowable state to a solid state in which particle contact forces span the system.
The exact nature of jamming is sensitive to particle-level details,
but, as we will see below, it bears some features of a continuous transition. 
We define a dense suspension as one
for which the macroscopic behaviour is governed
by physics associated with the proximity to this jamming transition.

\smallskip \noindent
\textbf{\emph{Out-of-equilibrium.}}
For the particle size range of interest here, 
the restoring force towards equilibrium,
Brownian motion,
may act over timescales comparable to, or longer than, a typical observation period.
This can be quantified by
the Stokes--Einstein relation for the diffusion coefficient
$\mathcal{D} \equiv k_{\mathrm{B}} T/6\pi\fluidvisc a$
(with $k_\mathrm{B}$ the Boltzmann constant and $T$ the temperature),
leading to a diffusive timescale $a^2/\mathcal{D} = 6\pi\fluidvisc a^3/k_{\mathrm{B}} T$.
For a small molecule this is $\approx \!\SI{e-10}{\second}$
and the local state is always very near equilibrium.
Meanwhile for a particle of radius $a=\SI{1}{\micro\meter}$ in water 
at room temperature it is $\approx \!\SI{1}{\second}$.
The $a^3$ dependence ensures that in many practical conditions,
where typical timescales are indeed of order \SI{1}{\second}, 
Brownian forces can be neglected when $a > \SI{1}{\micro\meter}$. 
Suspensions of such particles\,---\,these are called non-Brownian, 
athermal, or granular suspensions
and will be the focus of much of this article%
\,---\,%
are thus practically always out of equilibrium.
The exploration of configuration space therefore occurs not by thermal motion
but only as a result of external driving.
Consequently there is strong history dependence, 
and statements about macroscopic phenomena must be associated with 
a characterization of the prior strain protocol. 
Without this, there is no reference (equilibrium-like) state.

This out-of-equilibrium nature leads to non-Newtonian macroscopic stresses.
For simple fluids, which are Newtonian, normal and shear stresses 
are dominated by distinct physical mechanisms:
the former come from the stiffness of molecular interactions; 
the latter come from molecular diffusion inducing momentum diffusion. %
\begin{marginnote}[]
\entry{Stress}{
tensorial quantity $\bm{\Sigma}$ describing the force per unit area acting on the suspension. We write a general shear stress as $\sigma$ and pressure as 
$p \equiv -\mathrm{Tr}(\bm{\Sigma})/3 $}
\entry{Strain rate}{tensorial quantity $\tensor{E}$ defined 
as the symmetric part of a velocity gradient tensor 
$\nabla \bm{u}$: 
$\tensor{E} \equiv (\nabla \bm{u} + \nabla \bm{u}^T)/2$.
We write a general scalar strain rate $\gd$.
} 
\entry{Viscosity}{%
Relation between $\tensor{E}$ and the deviatoric part of $\bm{\Sigma}$. 
For a Newtonian fluid at hydrostatic pressure $p$, the viscosity $\eta$ 
is a (single) scalar parameter, 
reflecting the material character: 
$\bm{\Sigma} = - p \bm{I} + 2 \eta \tensor{E}$}
\end{marginnote}%
For suspensions, when describing the stress $\bm{\Sigma}$, one can distinguish the fluid stress $\bm{\Sigma}_\mathrm{f}$ coming from the solvent (often a simple fluid for which the above considerations hold)
from the particle stress $\bm{\Sigma}_\mathrm{p}$ due to mechanical forces transmitted through the solid phase. 
With diffusion suppressed (or at least as slow as the shear itself),
both normal and shear particle stresses originate in particles' inability
to follow the fluid streamlines
and their tendency to adopt a shear-induced structure,~\figref{figure2}\,\figpanel{h}.
The particle (and, hence, the total) stress components
are thus not independent of each other as they are for a simple liquid.

These distinguishing features make dense suspensions not amenable to an equilibrium description.
We therefore proceed by first giving an overview of the out-of-equilibrium phase 
transitions necessary to understand their behavior.

\section{OUT-OF-EQUILIBRIUM PHASE BEHAVIOUR} 
\label{sec:noneq}

Borrowing concepts from dry granular physics,
we first address 
the phase transition of 
major relevance
to much of the discussion that will follow on non-Brownian suspensions,
jamming.
We then give brief overviews of 
two transitions associated with Brownian suspensions (the glass transition), and suspensions of attractive particles,
acknowledging that separate reviews should be consulted for these important topics~\cite{berthier2011theoretical,zaccarelli2007colloidal}.

\subsection{Jamming transition}
\label{sec:jamming}

We consider
non-Brownian,
neutrally buoyant,
repulsive particle suspensions,
which generically transition from being 
flowable at low $\phi$ to being solid (and having a yield stress) at large $\phi$.
These limits are separated by a jamming transition, occurring at a critical volume fraction $\phij$ ($\approx 0.64$ for monodisperse frictionless spheres)
beyond which all motion is blocked due to widespread particle-particle contacts (as sketched in \figref{figure2}\,\figpanel{e}).
Above $\phij$, a (quasi-static) strain can only occur 
by deforming the particles,
which is not possible when they are rigid.
Jamming is not specific to suspensions: 
it occurs in dry granular matter, emulsions and other amorphous materials~\cite{liu_jamming_1998}.
It shares some features with equilibrium continuous phase transitions,
especially diverging length scales, though other quantities, such as the number of contacts,
may (dependent upon the protocol c.f.\ the preparation dependence mentioned above) 
vary discontinuously across the transition~\cite{liu2010jamming}.
 

The location of the transition, i.e.\ the value of $\phij$, is sensitive to many microscopic details
and also depends on the history of the sample%
~\cite{torquato_is_2000}.
Most importantly, inter-particle friction (quantified with dimensionless coefficients), which constrains rotational as well as translational particle motion, strongly affects jamming. 
All friction modes 
(sliding, rolling and twisting) contribute to a decrease in $\phij$~\cite{santos2020granular}, as does increasing the value of the friction coefficients.
It is particle-shape~\cite{donev_improving_2004,hsiao2017rheological} and size-distribution dependent:
broadening the particle size distribution from monodisperse to polydisperse
leads to a larger $\phij$~\cite{shapiro_random_1992,hopkinsDisorderedStrictlyJammed2013}.


These observations can be understood in a unified way by considering the appearance of isostaticity.
The suspension jams when just enough particle contacts form to constrain
all degrees of freedom in the system~\cite{alexander_amorphous_1998}.
This happens when the average number of contacts per particle $z$ reaches a critical isostatic value $z_\mathrm{c}$.
In general the value of $z_\mathrm{c}$ can be inferred 
from apparently simple Maxwellian constraint counting arguments~\cite{maxwell1864on}, 
though this rapidly becomes involved for
cases more complicated than frictionless spheres,
as each contact may constrain multiple degrees of freedom~\cite{shundyak_force_2007}.
Below jamming, i.e.\ $\delta z \equiv z_\mathrm{c} - z > 0$, the system is underconstrained,
and there are $N \delta z/2$ floppy modes in a system of $N$ particles.
These modes represent collective degrees of freedom along which the system can be deformed without elastic cost.
Floppy modes are spatially extended; due to the crowded conditions, moving a particle requires cooperative motion of other particles in the vicinity.


Jamming
has some features consistent with
a continuous transition.
Structural and mechanical properties behave singularly as a function of the distance to jamming $\delta \phi \equiv \phij -\phi$ 
(or equivalently as a function of $\delta z$),
and the associated exponents 
are often nontrivial~\cite{liu_jamming_2011,degiuli_unified_2015}.
In particular, several lengthscales diverge at the transition. 
Their identification and role is slowly emerging, with frictionless systems being better understood.
The most relevant for this review is the correlation length of the velocity field~\cite{olsson_critical_2007} 
(or the non-affine velocity field in the case of a sheared system),
quantifying the cooperativity of motion in crowded conditions.
Velocity correlations decay to zero on a typical length scale $l \sim \delta \phi^{-\lambda}$,
with numerical results for frictionless spheres giving
$\lambda\approx \numrange{0.6}{1}$~\cite{heussinger_jamming_2009}.
They can be quantitatively linked to the statistics of floppy modes, and scaling theory predicts that $l \sim 1/\sqrt{\delta z}$ and $\lambda \approx 0.43$~\cite{degiuli_unified_2015}.


This diverging length scale is central to the rheology close to jamming,
and is argued to be the origin of a divergence in the suspension viscosity $\suspvisc$, occurring at $\phij$~\cite{heussinger_jamming_2009}.
The physical picture is that correlated motion leads to non-affine velocities via a ``lever'' effect, which results in enhanced viscous dissipation~\cite{andreotti_shear_2012}.
The exact scaling relation (if any) between $l$ and $\eta$ is obscure, however, even for frictionless spheres.
In a caricature of correlated motion where particles are grouped 
in clusters of size $l$ rotating as rigid bodies~\cite{heussingerShearThickeningGranular2013}, 
equating for a cluster the injected power $\propto \eta_{\mathrm{f}} \dot\gamma^2 l^d$ to 
the dissipated power $\propto n \fluidvisc \dot\gamma^2 l^{d+2}$ (with $n$ the particle number density), 
one finds that $\eta \propto l^2$.
This is certainly too simplistic, however, as reaching $\phij$ from below the zero-shear viscosity (i.e.\ measured with infinitesimal shear stress) indeed diverges as  
$\eta \sim (\phij - \phi)^{-\beta}$, but with $\beta$ significantly larger than $2\lambda$, 
as we will see in \Secref{sec:rheology}.
Indeed, while eddies following approximately a rigid-body motion have a size of order $l$ and lead to the dominant decay of velocity correlations, there may be a secondary decay on a much longer length scale due to flow-induced contact anisotropy~\cite{during_length_2014}.
The associated correlation length, diverging faster than $l$ at jamming, would account for the viscosity divergence.


The jamming transition is thus key to understanding the rheology of dense suspensions and will form a central part of our discussion in what follows, but it is not the only relevant transition.

\subsection{Glass transition}

Brownian systems similarly undergo a transition from a flowable to a non-flowable state when increasing $\phi$.
This is the colloidal glass transition, occurring at $\phig$\,\cite{pusey1986phase}, 
above which particles are trapped in cages formed by their neighbors.
The glass transition is distinct from jamming~\cite{ikeda_unified_2012},
and we refer to other reviews for details about its phenomenology~\cite{hunter_physics_2012}.
The most relevant aspects for Brownian suspensions in this review are that
(i) 
the transition occurs before frictionless jamming, $\phig<\phij$ (for low polydispersity rigid spheres $\phig \approx 0.58$), 
(ii) below $\phig$ the zero-shear viscosity diverges in a super-exponential fashion, 
typically fitted with $\eta \sim \exp\bigl\{A/(\phig-\phi)^{\delta}\bigr\}$
(to be contrasted with the algebraic divergence at jamming),
and (iii) at $\phig$ a finite yield stress proportional to $k_\mathrm{B} T$~\cite{ikeda_unified_2012} 
appears discontinuously (whereas at jamming the yield stress appears continuously).

\subsection{A note about attractive interactions}

In the presence of attractive interactions,
particles can stick together into clusters that eventually span the system forming an elastic network
with a yield stress\,\cite{trappe2001jamming}. 
For Brownian systems, competition between $k_\mathrm{B}T$ and 
attraction leads to complex $\phi$-dependence\,\cite{Larson_1999}.
The resulting colloidal glass and gelation behaviour is an extensive and challenging topic and is reviewed elsewhere, 
see e.g.\ \cite{zaccarelli2007colloidal}.
Meanwhile in non-Brownian systems the role of attraction near jamming
is still emerging~\cite{tighe2018}.

\medskip

In what follows, we replace our labels for
$\phij$ and $\phig$ with a more general, stress-dependent limiting volume fraction $\phim(\sigma)$, representing the value of $\phi$ at which the suspension viscosity $\suspvisc$ diverges.
This is often still a \emph{jamming} point, but in the presence of Brownian or attractive forces the divergence may have contributions from multiple physics.

\section{RHEOLOGY AND MICROSCOPIC PHYSICS}
\label{sec:rheology}

\begin{figure}
\includegraphics[trim = 0mm 37mm 0mm 0mm, clip,width=1\textwidth,page=3]{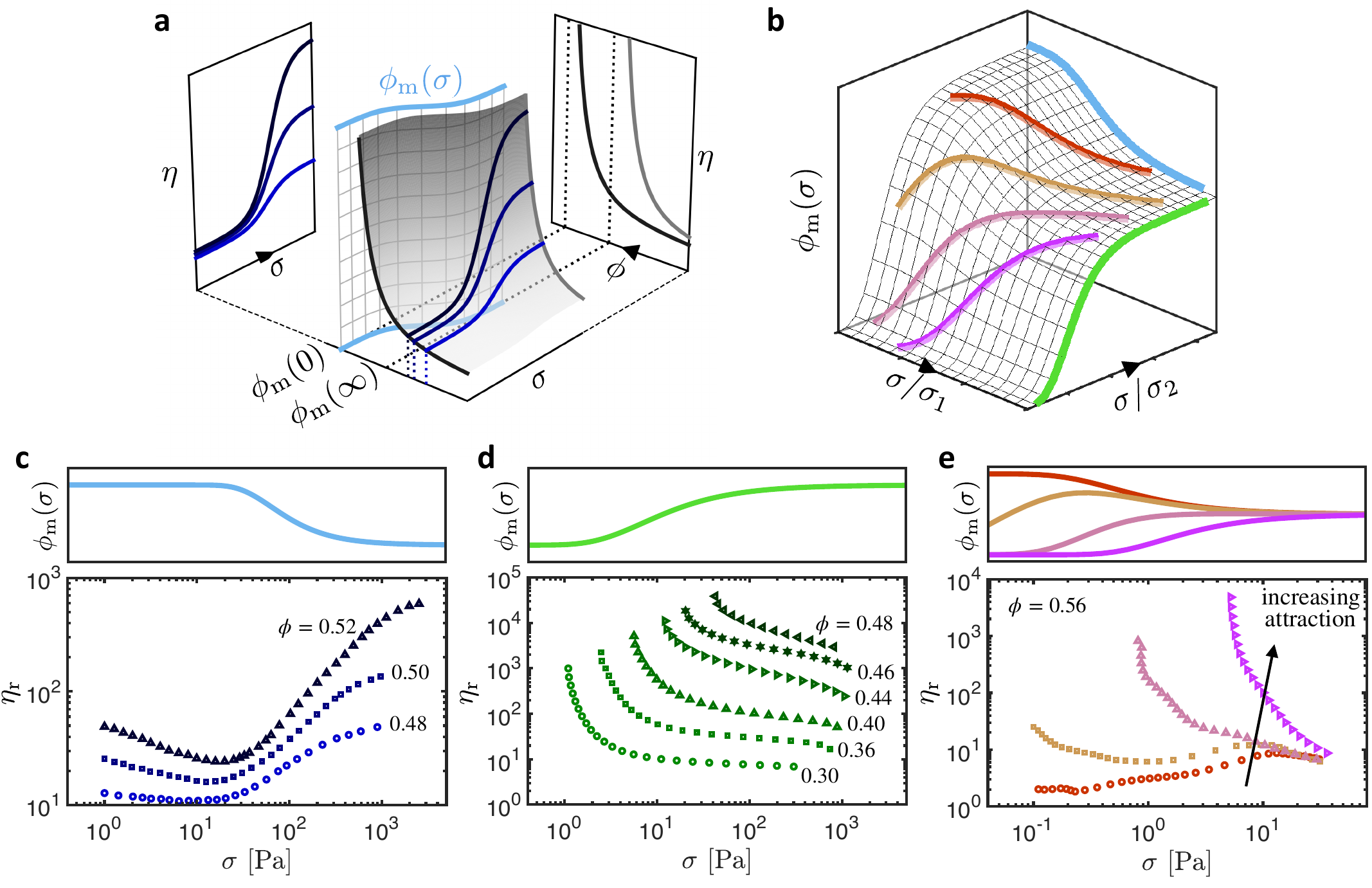}
\caption{%
Dense suspension rheology, showing schematics (\figpanel{a}, \figpanel{b} 
and top panels of \figpanel{c}, \figpanel{d}, \figpanel{e})
and experimental data (bottom panels of \figpanel{c}, \figpanel{d}, \figpanel{e}, 
with $\eta_{\mathrm{r}} \equiv \suspvisc/\fluidvisc$).
\figpanel{a}~%
Schematic relating the viscosity $\eta(\sigma,\phi)$ to $\phim(\sigma)$;
\figpanel{b}~%
schematic showing qualitative $\phim(\sigma)$ behaviour in the presence of one (blue and green) and two (red to magenta) characteristic stress scales, $\sigma_1$ and $\sigma_2$.
\figpanel{c}~%
Top: Schematic showing decreasing $\phim(\sigma)$;
Bottom: Experimental data showing shear thickening in a suspension of $a \approx \SI{260}{\nano\meter}$
silica particles in a polyethylene glycol solvent at 
three different $\phi$~\cite{cwalina2014material}.
Figure adapted with permission from Cwalina and Wagner,
Journal of Rheology 58(4):949--967 (2014);
copyright (2014) The Society of Rheology;
\figpanel{d}~%
Top: Schematic showing increasing $\phim(\sigma)$;
Bottom: Experimental data showing shear thinning in a suspension of
$a \approx \SI{4}{\micro\meter}$ ground calcium carbonate particles in a glycerol-water mixture at 
a range of $\phi$~\cite{richards2020turning}.
Figure adapted with permission from Richards et al,
Rheologica Acta 60:97--106 (2021);
copyright (2021) Rheologica Acta;
\figpanel{e}~%
Top: Schematic showing $\phim(\sigma)$ for (top to bottom) increasing attraction.
Bottom:
Experimental data for 
a suspension of 
$a \approx \SI{90}{\micro\meter}$
soda-lime glass spheres in mineral oil at $\phi=0.56$.
Shown (bottom to top) are increasing applied electric field strengths, controlling attractive interactions~\cite{brown2010generality}.
Figure adapted with permission from Brown et al,
Nature Materials 9:220--224 (2010);
copyright (2010) Springer Nature.}
\label{figure3}
\end{figure}

Having introduced the transitions that occur as $\phi$ approaches $\phim$, we now discuss how the proximity to them is relevant in determining the macroscopic response,
in particular the rheology under uniform flow (conditions in which the strain rate is spatially uniform).
In general the suspension viscosity $\suspvisc$ diverges at $\phi=\phim$,
and complex rheology can be understood as stress- or rate-controlled
changes to the dominant microphysics,
in many circumstances leading to stress dependence in $\phim$ itself.
Such a scheme is sketched in \figref{figure3}\,\figpanel{a}.
Here we illustrate how a monotonic $\phim(\sigma)$ (the physics of which we consider below) leads naturally to
distinct stress-dependent viscosity divergences
(whose form may follow e.g. $\suspvisc \sim [\phim(\sigma)-\phi]^{-\beta}$) 
and thus to rate-dependent rheology $\suspvisc(\sigma,\phi)$.
Rather than following a top-down approach to addressing macroscopic phenomenology,
we instead consider systematically the microscale physics sketched in \figref{figure2}.
Our focus is not on the origin of microscopic forces \textit{per se} but rather 
on their effect on the rheology.
For experimental and computational methodologies\,---\,rheometry\,---\,we refer the reader to other reviews 
(see e.g.\ \cite{coussot2016rheophysics,maxey2017simulation}).
We take as our starting point the idealised case of \emph{rate-independent} suspensions (for which $\phim$ is independent of $\sigma$),
before addressing rheology that arises in more complex cases, systematically examining physics that are
\emph{one dimensionless parameter} away from the rate-independent case.

\subsection{A reference point: rate-independent suspensions}
\label{sec:rate_indep_suspensions}

Consider a dense suspension of non-Brownian, neutrally buoyant, rigid particles in a Newtonian fluid, 
subject to a steady deformation slow enough that fluid and particle inertia can be neglected.
In practice this can be approximated with a suspension of $\approx\!\SI{10}{\micro\meter}$ glass particles in an appropriate solvent under typical rheometric conditions.
Such a system plays a special model role that has proven to be hugely influential in understanding 
many other classes of suspensions,
as we will see in what follows.

\paragraph*{Dimensional analysis.}
The microscopic quantities involved in the problem are the typical particle size $a$ and density $\partdens$, and the fluid viscosity $\fluidvisc$
and density $\fluiddens$ (with $\partdens = \fluiddens$).
For rigid particles there is no microscopic energy or force scale.
The macroscopic quantities are the control volume size $L$
(this is taken to be sufficiently larger than $a$
so that we consider bulk phenomena)
and the solid fraction $\phi$ therein.
Finally the deformation is characterized by its rate $\gd$, the applied stress $\sigma$, 
and the time for which it was applied, $t$.
(This notation is conventional for simple shear,
but the following argument applies generally.)
Buckingham~$\pi$ implies that there are 4 dimensionless numbers besides $\phi$. 
Appropriate choices are the Stokes number $\StNum \equiv \partdens \gd a^2/\fluidvisc$, 
the Reynolds number $\ReNum \equiv \fluiddens \gd L^2/\fluidvisc$,
the relative suspension viscosity $\eta_\mathrm{r} \equiv \suspvisc/ \fluidvisc = \sigma/\fluidvisc \gd$, and the strain $\gamma = \gd t$. 

As stated earlier, we take as our reference point suspensions for which particle and fluid inertia can be neglected, 
corresponding, respectively, to $\StNum = 0$ and $\ReNum = 0$.
We then conclude that $\eta_\mathrm{r}$ must be a function of $\phi$ and $\gamma$ only. 
This is a strong statement: it implies that the stress is linear in the deformation rate, 
as for a Newtonian fluid, 
and that the proportionality factor, the suspension viscosity $\eta$, only depends on $\phi$ and $\gamma$.
Furthermore, in steady state (the $\gamma \to \infty$ limit, in practice typically
achieved for $\gamma = \mathcal{O}(1)$), $\eta_\mathrm{r}$ is function of $\phi$ only.

This analysis (which follows Krieger \cite{krieger_dimensional_1963}) can be extended to any component of the stress tensor, 
though these do not each necessarily share the same parity with $\gd$. 
While shear stresses (or more generally the dissipative part of the stress tensor~\cite{Giusteri_2018}) 
are linear in $\gd$, normal stresses, which for a rate-independent suspension 
are always negative (pushing outward), are linear in $|\gd|$.
Depending on the system, the list of dimensionless numbers will be augmented, most notably by the particle friction coefficient(s).
Recent measurements show that friction is indeed a generic feature of direct contacts among $\si{\micro\metre}$ (and larger) sized particles%
~\cite{fernandez_microscopic_2013,comtet_pairwise_2017}, 
though the non-dimensionality of the coefficient ensures (assuming friction is Coulombic) 
that this alone does not introduce rate dependence.
This \emph{quasi-Newtonian} result will form the basis for our discussion of more complex, rate-dependent, suspensions in what follows.

\paragraph*{Rheology under steady flow.}

Under a steady imposed deformation rate tensor, rate-independent suspensions can thus be characterized 
by a set of material functions\,\cite{Giusteri_2018}
all depending on $\phi$ only.
As expected from the discussion in \Secref{sec:jamming},
these are all singular at jamming, 
typically behaving asymptotically as $(\phim-\phi)^{-\beta}$~\cite{boyer2011unifying}.

Here $\phim$ typically represents a frictional jamming point (but see \Secref{sec:repulsive}
 below).
Many experiments~\cite{boyer2011unifying,ovarlez_local_2006} as well as simulations 
of spherical particle suspensions~\cite{andreotti_shear_2012,ness_flow_2015} are consistent with $\beta\approx 2$, 
although quite different values have also been reported, 
with a possible dependence on the sliding friction coefficient $\mu_\mathrm{p}$%
\,\cite{
zarraga_characterization_2000,
gallier_fictitious_2014,
trulsson_effect_2017}.
In a simplified model of frictionless sphere flow, for which the exponent can be determined with higher precision thanks 
to efficient numerical methods and to scaling theory linking $\beta$ to other independently measurable exponents, 
one finds $\beta\approx 2.8$~\cite{lerner_unified_2012}.
For long rodlike particles, it has been experimentally measured as $\beta \approx 1$%
~\cite{tapia_rheology_2017}.

This rheology can also be formulated from an alternative (but equivalent), particle-pressure-imposed, perspective.
Boyer et al~\cite{boyer2011unifying} demonstrated (and verified experimentally) that in simple shear 
the stress anisotropy,
or macroscopic friction coefficient,
$\mu$ (which in simple shear is $\mu\equiv \sigma/P_\mathrm{p}$, with $P_\mathrm{p}$ a measure of the particle pressure) 
and $\phi$ are functions of the so-called viscous number $J \equiv \fluidvisc \gd/P_\mathrm{p}$ only.
Knowing $\mu(J)$ and $\phi(J)$,
we can use that $\suspvisc/\fluidvisc = \mu(J)/J$ and invert the $\phi(J)$ relation to recover 
the volume-imposed rheology,
that is, a rate-independent viscosity depending on $\phi$ only.
Here jamming corresponds to $J\to 0$. 
In this limit, $\phi(J) \sim \phim - aJ^{1/\beta}$ and $\mu(J) \sim \mu^{\ast} + bJ^{\xi}$%
~\cite{boyer2011unifying}, 
such that $\suspvisc \sim (\phim - \phi)^{-\beta}$.
Importantly, $\mu^{\ast}$ is finite, implying that particle normal and shear stresses share the same 
asymptotics close to jamming.
As with $\phim$, $\mu^{\ast}$ depends on microscopic details. 
For frictionless ($\mu_{\mathrm{p}}=0$) spheres,
$\mu^{\ast} \approx 0.1$~\cite{dacruz2006rheophysics},
whereas for $\mu_\mathrm{p} \gtrsim 0.1$,  
$\mu^{\ast} \approx $ \numrange{0.3}{0.4}~\cite{chevremont_quantitative_2019}. 
Interestingly, the large friction limits for $\mu^{\ast}$ and $\phim$ occur at quite distinct values of $\mu_\mathrm{p}$~\cite{chevremont_quantitative_2019}.
%

\paragraph*{Microstructure evolution and flow history dependence.}
For rate-independent suspensions,
statements about the macroscopic rheology must be associated with a description of their strain history (though the \emph{rate} at which this history was explored is unimportant).
An initial transient of a few strain units ($\gamma=\mathcal{O}(1)$)~\cite{gadala1980shear}
is generally observed upon flow start-up,
after which the memory of initial conditions is lost
and particles have, thanks to shear-induced diffusion~\cite{leighton1987measurement},
sampled a statistically representative part of the configuration space.
Steady states (with $\gamma\gg1$) are thus unambiguously defined, independent of the prior sample history, and are the usual reference points for characterising rate-independent rheology.

Dense suspensions usually exhibit microstructural anisotropy, details of which depend on the deformation being applied, see e.g.~\figref{figure2}\,\figpanel{h}.
This is often characterized at the level of the statistics of some particle interaction director $\bm{n}$, usually via its averaged second moment, referred to as the structure, texture or fabric tensor $\langle \bm{n}\bm{n}\rangle$ (see also \Secref{sec:const}).
During steady state flow, this tensor is often found to be aligned with the deformation rate tensor~\cite{parsi1987fore,brady1997microstructure}.
The microstructural anisotropy is mirrored at the macroscopic level by the stress anisotropy $\mu$, although their exact relation remains elusive. 
For frictionless spheres near jamming, microstructural anisotropies bias the floppy mode statistics in favor of flow-resisting modes, leading to larger $\mu$~\cite{degiuli_unified_2015}.
Meanwhile, first and second normal stress differences quantify, respectively, 
stress anisotropy in and out of the flow plane~\cite{Giusteri_2018,zarraga_characterization_2000, Seto_2018}.

Together, the finite transient in the strain~\cite{gadala1980shear,han2018shear}
and the shear-induced structure give rise to an anisotropic stress response: 
the viscosity of a rate-independent suspension is largest 
when the strain rate and microstructure tensors align.
This is readily illustrated under unsteady flow conditions such as shear reversal~\cite{gadala1980shear} and oscillatory shear~\cite{blancTunableFallVelocity2014}.
When the direction of the applied flow changes more rapidly than a steady microstructure 
can establish (i.e.\ accumulating strains $\gamma < 1$ before reversal),
the result is a systematically lower viscosity than in steady shear.
An extreme case of this is the remarkable phenomenon of shear jamming~\cite{Cates_1998a,Seto_2019a},
in which a suspension is jammed under a previously applied deformation, but not jammed with respect to others.
Separately, the behaviour under repeated reversals
reveals,
for sufficiently small strain amplitude,
contact-free states~\cite{pine2005chaos}, indicating the presence of a nearby absorbing state~\cite{tjhung2015hyperuniform}.

\subsection{Particle Inertia}
\label{sec:inertia}

At the upper end of our size range ($a\approx\SI{1}{\milli\meter}$), particle inertia can be significant, i.e.\ $\StNum = \mathcal{O}(1)$, even at modest $\dot\gamma$.
Dimensional analysis then implies an inertial stress that scales with $\partdens a^2\gd^2 g(\phi)$, 
the so-called Bagnold scaling~\cite{bagnold_experiments_1954},
consistent with experimental~\cite{savage_shear_1983,madraki2020shear}
and numerical~\cite{%
verberg_rheology_2006, kulkarni_suspension_2008, trulsson_transition_2012}
observation.
Assuming that viscous ($\propto\!\gd$) and inertial ($\propto\!\gd^2$) stresses are additive, 
one concludes that the former will dominate at low shear rates and the latter at high rates, 
with a crossover occurring at $\gd_\mathrm{in}$.
Above $\gd_\mathrm{in}$, the rate-independence described above thus gives way to a stress that is quadratic in $\gd$
(leading to $\suspvisc \propto\gd$),
a form of continuous shear thickening.

As with rate-independent flow, inertial rheology can be expressed in constant pressure terms.
Whereas $J$ characterises the rheology under purely viscous conditions, the appropriate dimensionless shear rate under purely inertial conditions is the inertial number 
$I \equiv a \gd \sqrt{\partdens /P_\mathrm{p}}$ 
with $\mu$ and $\phi$ now functions of $I$ only~\cite{jop2006constitutive,lemaitre_what_2009}.
Generally, a viscous regime is expected for $\StNum = I^2/J\ll 1$, and an inertial one for $\StNum \gg 1$~\cite{trulsson_transition_2012}.

A central and unresolved question is that of the dependence (if any) of $\gd_\mathrm{in}$ on $\phi$ 
when approaching $\phim$.
The inertial and viscous stresses are expected to scale, respectively,
as $\gd^2 (\phim -\phi)^{-\beta_\mathrm{i}}$
and
$\gd (\phim -\phi)^{-\beta_\mathrm{v}}$~\cite{lemaitre_what_2009},
with $\phim$ usually considered the same for both
since inertia shouldn't affect the isostaticity condition (though a stringent test of this is absent).
%
%
Equating the stress contributions at the crossover leads to $\gd_\mathrm{in} \sim (\phim -\phi)^{-\beta_\mathrm{v} + \beta_\mathrm{i}}$.
Several numerical works report $\beta_\mathrm{v}=\beta_\mathrm{i}=2$, 
or equivalently that $\mu$ and $\phi$ are functions of a composite number $K =J+\alpha I^2$ with $\phim - \phi \sim K^{1/2}$~\cite{ness_flow_2015,trulsson_transition_2012}. 
Meanwhile others~\cite{otsuki_universal_2009,otsuki_critical_2011,vagberg_critical_2016}, as well as experiments~\cite{fall2010shear}, report $\beta_\mathrm{i} > \beta_\mathrm{v}$ (usually $\beta_\mathrm{i} \approx 2 \beta_\mathrm{v} \approx \numrange{4}{5}$, compatible with scaling theory for flow near jamming~\cite{degiuli_unified_2015,trulsson_effect_2017},
which predicts $\beta_\mathrm{i} = 2 \beta_\mathrm{v} \approx 5.7$),
implying that inertia is relevant down to a Stokes number vanishing with proximity to jamming. 
However, the stress below which inertia can be neglected is $\suspvisc \gd_\mathrm{in} \sim (\phim -\phi)^{-2\beta_\mathrm{v} + \beta_\mathrm{i}}$, which remains finite at jamming if $\beta_\mathrm{i} = 2 \beta_\mathrm{v}$.

\subsection{Brownian motion}

Following the \emph{rule of thumb} in \Secref{sec:intro},
Brownian forces become important for particles with $a< \SI{1}{\micro\metre}$,
leading to the characteristic stress scale $k_{\mathrm{B}} T/a^3$~\cite{batchelorEffectBrownianMotion1977} becoming relevant.
Thermal fluctuations
allow such Brownian particles to explore configuration space in the absence of external driving,
so that an underlying `equilibrium' phase diagram might be defined~\cite{pusey1986phase}
(though often in practice the system will not reach equilibrium over typical observable time scales).

Under flow, the competition between characteristic timescales for particle diffusion, which acts to
re-disperse any shear-induced microstructure
and restore equilibrium, and convection, which drives the system out-of-equilibrium, is quantified by a dimensionless shear rate,
the P\'{e}clet number $\PeNum \equiv 6\pi\fluidvisc a^3\gd/k_\mathrm{B} T$.
This controls a crossover between the relative importance of the Brownian and viscous stresses.
The former is
sublinear in $\dot\gamma$
due to $\PeNum$ dependence of the microstructure~\cite{fossStructureDiffusionRheology2000},
and thus Brownian suspensions typically shear thin~\cite{woods1970rheological,de1985hard}.
There may be a high shear (in practice this means $\PeNum \gg 1$) viscosity plateau,
where rate-independence is recovered (unless other physics intervene, see below).


An alternative perspective may be to consider Brownian shear thinning
as the consequence of a $\PeNum$-controlled change in the limiting volume fraction~$\phim$~\cite{ikeda2012unified}.
It is established that the viscosities
of Brownian and non-Brownian suspensions
diverge at 
distinct points~\cite{mewis2012colloidal,de1985hard},
and these may be associated with 
the glass $\phi_\mathrm{G}$ and jamming $\phij$ transitions respectively
(recalling that $\phij > \phi_{\mathrm{G}}$).
Hence, increasing $\PeNum$ at a given $\phi$ leads to $\phim$ increasing from $\phi_\mathrm{G}$ to $\phij$,
resulting in shear thinning
(or indeed yielding if $\phi_{\mathrm{G}} <\phi< \phij$~\cite{ikeda2012unified}).

\subsection{Repulsive interactions}
\label{sec:repulsive}

In reality particles usually have repulsive interactions acting over a finite range,
stabilising them against clustering, inhibiting contact formation, 
and setting a force scale that is absent in the rigid particles considered so far.
These interactions may originate in electrostatics, polymer coatings,
or from other physics (Brownian motion may provide an effective repulsion~\cite{trulsson_athermal_2014}).
Importantly,
a characteristic repulsive stress $F_{\mathrm{r}}/a^2$ competes with the viscous one $\propto \! \fluidvisc \gd$,
and a dimensionless shear rate can be defined e.g.\ as $\dot{\gamma}/(F_{\mathrm{r}}/\fluidvisc a^2)$.
Details, including the range of the repulsive force,
control how this quantity
governs the effective proximity to $\phim$
and the consequent rate-dependence that emerges.

As mentioned in \Secref{sec:rate_indep_suspensions},
direct contacts among $\si{\micro\metre}$ (and larger) sized particles are typically frictional.
The short range repulsive interactions described above act to inhibit the formation of such contacts between particles, maintaining lubrication layers and thus limiting (in a stress-controlled manner) the role played by friction.
The implications of this are profound and far-reaching (for a review, see~\cite{morris2020shear}).
As the applied stress $\sigma$ is increased relative to an \emph{onset} stress $\sigma^{\ast}\propto F_{\mathrm{r}}/a^2$, the repulsive barrier is overcome and particle contacts increasingly transition from the lubricated to the frictional contact state.
(The $1/a^2$ dependence ensures that particles larger than a few microns are persistently frictional.)
With friction present, fewer contacts are required for mechanical equilibrium i.e.\ $z_c$ is reduced~\cite{shundyak_force_2007}.
$\phim$ will consequently \emph{reduce},
from a frictionless limit $\phim^{(0)}\equiv\phim(\sigma/\sigma^{\ast}\to0)$ 
to a lower, 
frictional value $\phim^{(1)}\equiv\phim(\sigma/\sigma^{\ast} \to\infty)$, as illustrated by the solid light blue line in~\figref{figure3}.
Under fixed $\phi$, therefore, increasing $\sigma$ brings the suspension closer 
to jamming,
by reducing $\phim(\sigma)-\phi$. 
This ``frictional transition'' mechanism (encoded in a popular constitutive model~\cite{wyart2014discontinuous}, described later) 
provides a generic route to shear thickening, 
distinct from the  inertial Bagnoldian rheology described above~\cite{ness2016shear}.
(When present, inertia may nonetheless contribute to mediation of particle contacts, thereby playing a role in a dynamic transition mechanism and enabling repulsionless frictional shear thickening~\cite{otsuki_critical_2011,fernandez_microscopic_2013}.)
It was proposed based on theory and simulation~\cite{seto2013discontinuous,wyart2014discontinuous,heussingerShearThickeningGranular2013} 
and has been confirmed by a series of experiments, see~\cite{fernandez_microscopic_2013,guy2015towards,lin2015hydrodynamic} among many others.
Example experimental flow curves $\suspvisc(\sigma,\phi)$ are shown in 
\figref{figure3}\,\figpanel{c}~\cite{cwalina2014material}.


The extent of shear thickening increases as $\phi$ approaches $\phim^{(1)}$.
Close to this value, the viscosity increase becomes discontinuous (indeed theory predicts that $\gd$ becomes non-monotonic in $\sigma$),
while above it flow stops at stresses of order $\sigma^{\ast}$.
Shear thickening by this mechanism bears some hallmarks of a phase transition:
increasing correlation lengths;
stress fluctuations~\cite{rathee2017localized};
separation into lubricated and frictional states~\cite{fall2009yield,hermes2016unsteady,chacko2018dynamic}.

The above discussion applies generally to repulsive particles,
while increasing the \emph{range} of repulsion can introduce a separate effect.
There exists a characteristic particle separation $h$ 
for which the repulsive force balances the typical viscous force 
$\propto \! \fluidvisc \gd a^2$.
Here, $a_\mathrm{eff} \equiv (a+h/2)$ acts as the \emph{effective} radius of a larger, soft particle,
leading to an effective $\phi_\mathrm{eff}>\phi$.
The viscosity, 
now set by e.g.\ $\suspvisc \sim(\phim-\phi_\mathrm{eff})^{-\beta}$, 
is thus enhanced by the presence of the repulsive force
(there may even be a finite yield stress if $ \phi_{\mathrm{eff}} > \phim $).
%
%
%
As $\gd$ increases,
the force balance (assuming the repulsive force increases with decreasing separation) dictates that $h$ decreases, as do $a_\mathrm{eff}$ and $\phi_\mathrm{eff}$ (though this must remain $\geq \phi$).
The system consequently moves further from $\phim$,
leading to reduced $\suspvisc$.
Low-$\gd$ shear thinning arising by this simple argument is a common feature of experimental rheology data~\cite{maranzano2001effects}.

\subsection{Attractive interactions}
\label{sec:attractive}

When present and sufficiently large,
van der Waals (or other, for instance depletion) forces
lead to attraction between particles. 
This introduces a competition between aggregation processes (sticking particles together and providing elasticity~\cite{rueb1997viscoelastic}) and shearing processes (breaking particles apart).
Steady shear generically breaks attractive bonds between particles, destroying larger flocs or clusters and leading to shear thinning~\cite{cross1965rheology,woutersen1991rheology,rueb1998rheology},
a typical characteristic of \textit{pastes}~\cite{roussel2010steady}.
For Brownian particles, 
aggregation processes are well-understood in the context of colloidal gelation~\cite{zaccarelli2007colloidal}
and their influence on the rheology is well-established~\cite{mewis2012colloidal}.

For non-Brownian particles, however, aggregation is instead driven by external forces, often shear itself~\cite{guery2006irreversible}.
There is some evidence that the same general picture of shear thinning due to breaking aggregates applies both for weakly attractive particles~\cite{snabre1996rheology} and rods~\cite{chaouche2001rheology},
and more strongly attractive systems~\cite{zhou1995yield,kurokawa2015avalanche}.
In the absence of shear, meanwhile,
numerical evidence~\cite{tighe2018} suggests that attraction reduces $\phim$ in non-Brownian, frictionless systems.
Recent experimental measurements support
the fact that weak attraction can
enhance particle contacts~\cite{park2019contact}
and even stabilise them against rolling~\cite{richards2020turning}.
The combination of attraction and restricted tangential motion, \emph{adhesion}
(or \emph{cohesion} if the particles are identical),
generates a yield stress at lower volume fractions still,
details of which are heavily protocol dependent~\cite{richards2020role}.
Shearing at sufficiently large $\sigma$ is argued to
break adhesive contacts and relieve tangential constraints on particles,
contrary to stress-induced friction which introduces sliding constraints as the stress is increased.
Under steady shear, therefore,
$\phim$ is reported to increase with $\sigma$ (solid green line in~\figref{figure3}),
providing evidence that the yielding and shear-thinning rheology of attractive, 
non-Brownian suspensions (an example experimental flow curve $\suspvisc(\sigma,\phi)$ is shown in~\figref{figure3}\,\figpanel{d}~\cite{richards2020turning}) 
might be characterised within the same framework of stress controlled changes to $\phim$~\cite{guy2018constraint} (see also~\cite{wildemuth1984viscosity}). 
It is not yet clear whether the few numerical works on attractive, frictional suspensions corroborate this~\cite{pednekar2017simulation,singh2019yielding},
in part due to the challenge of precisely defining particle contact models comprising attraction, rolling and sliding friction.

\subsection{More complex suspensions}

Most suspensions will, in practice, have more than one relevant stress scale,
with various physics from the Sections above being important under typical operating conditions.
For example, the combined effects of increasing and decreasing $\phim$ due to, respectively, 
an adhesive stress~\cite{richards2020turning} (or indeed a Brownian stress~\cite{laun1984rheological}) 
and a repulsive stress, can lead to non-monotonic rheology.
We show such an example in~\figref{figure3}\,\figpanel{e}, 
taking experimental data from~\cite{brown2010generality} (other examples include~\cite{gopalakrishnan2004effect}).
By tuning the relative importance of one of the dimensionless groups 
(in the example shown this is achieved by manipulating the attraction),
one tunes between shear thinning and thickening rheology.
This might be characterised as a transition, as a function of $\sigma$, across a multidimensional $\phim$ map, \figref{figure3}\,\figpanel{b}.
Here we plot such a surface (following~\cite{guy2018constraint}, see~\Secref{sec:const-ratedep}) comprising both $\phim$-reducing and $\phim$-increasing physics.
Besides the limiting cases depending on only one stress scale
(in blue is a typical shear-thickening suspension; in green is an adhesive, shear thinning one),
a more general scenario with two (or more) stress scales ($\sigma_1, \sigma_2, \dotsc$)
that lead to much richer $\phim(\sigma)$ may be useful to interpret the rheology of more complex suspensions.
For instance, \figrefp{figure3}{e}
links complex experimental rheological data to the putative $\phim(\sigma)$ map sketched in~\figrefp{figure3}{b}.
It is, however, still too early to assert this scenario, and more experimental, numerical, and theoretical works in this direction are needed.

Dense suspensions with a broad particle size distribution present more complexity still.
Although polydispersity is known to increase $\phim$ and consequently decrease the viscosity of rate-independent suspensions~\cite{shapiro_random_1992},
the effect of broad polydispersity on rate-dependent rheology is not well understood.
Whereas for roughly monodisperse suspensions
the microscopic stress scales described above map to macroscopic ones,
for polydisperse suspensions this is more involved.
In practice particle radii $a$ may span 6 orders of magnitude~\cite{roussel2010steady}.
In such cases it is not clear how to define bulk dimensionless control parameters, 
which have, for instance, $St\propto a^2$, $Pe \propto a^3$ and contact friction onset $\sigma^* \propto 1/a^2$.
Extending the descriptions above to such systems poses a challenge~\cite{guy_testing_2019},
as do numerous other sources of complexity (we list a few as Future Issues below).

\section{CONTINUUM MODELS AND FLUID MECHANICS}
\label{sec:const}

The above insights can be utilised to make predictions of suspension behaviour in practical scenarios.
Doing so requires continuum models,
giving the time evolution of the velocity field, driven by the stress field through the Cauchy equation,
and other macroscopically relevant fields
(the list of which is itself a modelling challenge), 
such as microstructure and volume fraction.
Such models must combine conservation laws with closures for the coupling terms relating the different fields.


We first address constitutive equations,
i.e.\ closures relating the stress tensor to the (history of) deformation.
These should describe both steady and transient phenomena,
the latter encoding the relative slowness (dictated by the shear rate)
of the microstructural evolution.
We limit the discussion to \emph{local} models,
acknowledging that non-local phenomena,
relatively well-studied in dry granular matter
(see e.g.\ \cite{bouzid2015non,kamrin2012nonlocal}),
likely play a role in suspensions near $\phim$.
We then give a brief introduction to two-phase continuum equations, 
necessary for describing flows involving migration,
relative motion between solid and liquid phases~(\figref{figure2}\,\figpanel{i}).
We conclude with a brief overview of some relevant aspects of the fluid mechanics of dense suspensions.

\subsection{Rate-independent microstructural constitutive models}

As briefly introduced in \secref{sec:rate_indep_suspensions}, the stress is strongly coupled to the anisotropy of the microstructure. 
The mathematical nature of the stress (being a \nth{2} order, symmetric tensor) naturally calls for a coupling with a suitably defined fabric tensor.
The main approach for rate-independent suspensions thus follows 
the anisotropic fluid theory of Hand~\cite{hand_theory_1962}.
This addresses the time evolution of a fabric tensor $\langle \bm{n} \bm{n} \rangle$  
under a velocity gradient tensor $\nabla \bm{u}$.
The director $\bm{n}$ represents, for instance, 
the orientation of particle contacts.
If $\bm{n}$ is not uniformly distributed, its typical orientation is given by the eigenvector of $\langle{\bm n}{\bm n}\rangle$ associated to the largest eigenvalue.
Such a theory then consists of an expression for the stress tensor $\bm{\Sigma}$ 
as a function of $\langle \bm{n} \bm{n} \rangle$ and the symmetric part of 
the deformation rate tensor 
$\tensor{E} \equiv (\nabla \bm{u} + \nabla \bm{u}^T)/2$, 
and a dynamics for $\langle \bm{n} \bm{n} \rangle$. 
The latter depends on $ \nabla \bm{u}$ and is constrained 
by frame indifference when particle inertia is negligible (see Sec. 4.3 in~\cite{Phan-Thien_2017}), and for a homogeneous system is:
\begin{equation}
\bm{\Sigma} = \bm{\Sigma}(\langle \bm{n} \bm{n}\rangle, \tensor{E}), 
\qquad \frac{D \langle \bm{n} \bm{n} \rangle}{D t} = \mathcal{F}\left(\langle  \bm{n} \bm{n} \rangle, 
\tensor{E} \right),
\end{equation}
with $D\langle \bm{n} \bm{n} \rangle/ Dt \equiv 
d\langle \bm{n}\bm{n} \rangle/d t + 
\langle \bm{n} \bm{n} \rangle \cdot \bm{\mathit{\Omega}} - \bm{\mathit{\Omega}}\cdot \langle \bm{n} \bm{n} \rangle$ the co-rotational (or Jaumann) derivative and $\bm{\mathit{\Omega}} \equiv (\nabla \bm{u} - \nabla \bm{u}^T)/2$.
There are two routes to such a theory.
The first is a phenomenological one, where one keeps the structural form of these equations 
as general as permitted by symmetries (e.g.\ \cite{goddard2006dissipative,chacko2018shear,ozenda2018new}), 
which keeps the theory flexible at the cost of having many free parameters and little understanding of their microscopic origin.
The other route is microstructural, with $\mathcal{F}\left(\langle \bm{n} \bm{n}\rangle, \nabla \bm{u} \right)$ derived from 
particle dynamics using simplifying assumptions%
~\cite{hinch_constitutive_1976,phan1995constitutive,gillissen2019constitutive}.
This limits the number of free parameters but involves (often uncontrolled) approximations,
in particular the closure of higher moments of $\bm{n}$, 
such as $\langle \bm{n} \bm{n} \bm{n} \bm{n}\rangle$ 
in terms of 
$\langle \bm{n} \bm{n}
\rangle$~\cite{szeri_new_1994,chacko2018shear}.
While phenomenological models can achieve quantitative agreement with experimental data regarding the transient behavior of the shear stress in simple shear%
~\cite{ozenda2018new}, 
their predictions for normal stresses (when tested) are usually poorer~\cite{stickel_application_2007,chacko2018shear}  
(although some aspects of the steady state behavior can be captured~\cite{ozenda_tensorial_2020}).


Recent works revisited the microstructural route based on the physical picture of a nearby jamming transition, 
giving a quantitative agreement with simulations for shear stresses and a qualitative one for normal stresses~\cite{gillissen2019constitutive}.
The key assumption is that due to the one-sidedness of rigid particle contacts (which dominate the rheology close to jamming), 
the microstructure reacts in a strikingly different manner to compressive and extensional strains.
This motivates a phenomenological theory where, defining a decomposition 
$\tensor{E} = \tensor{E}_\mathrm{c} + \tensor{E}_\mathrm{e}$ 
one gives distinct roles to the compressive $\tensor{E}_\mathrm{c}$ and extensional $\tensor{E}_\mathrm{e}$
strain rate tensors in the dynamics of $\langle \bm{n} \bm{n} \rangle$ 
and in the stress--structure relation~\cite{gillissen_modeling_2018}.
%


These rate-independent equations
may constitute a basis for the development of constitutive models for suspensions with more complex interactions, 
as we will discuss in the next subsection.
It should however be noted that so far, even for works following the microstructural route, 
the stress--structure relation is phenomenological, and not derived from the microscopic dynamics.
In particular, the viscosity divergence at jamming, when considered, is assumed to follow the algebraic form discussed in \Secref{sec:rheology}. 
Microscopically derived tensorial stress--structure relations are still lacking, 
even though real progress has been achieved for scalar relations 
in model suspensions~\cite{degiuli_unified_2015,trulsson_effect_2017}.

\subsection{Constitutive models for rate-dependent suspensions}

\label{sec:const-ratedep}

A typical approach in writing steady state rheological models is to define (for fixed $\phi$) 
limiting $\eta_0 \equiv \eta(\gd\to 0)$ and $\eta_\infty \equiv \eta(\gd\to\infty)$ 
(in principle these can be measured experimentally) 
and then specify a smooth transition between them as a function of, 
e.g.\ $\dot{\gamma}_{\mathrm{c}}/\dot{\gamma}$ or $\sigma_{\mathrm{c}} /\sigma$~\cite{carreau1972rheological}.
Such ``rheological interpolation'' models are widely used for the fitting of data~\cite{mewis2012colloidal},
and in some instances $\gd_{\mathrm{c}}$ (or $\sigma_{\mathrm{c}}$) 
may provide microstructural insight~\cite{cross1965rheology}.


As discussed in \Secref{sec:rheology}, the physics and rheology of several important classes of dense suspensions can be described as the result of an alteration of $\phim$ 
under flow due to the competition between applied stress and microscopic interparticle forces at work.
This basic mechanism can be formalized with rheological interpolation models
that \emph{do} have a microscopic grounding, as shown initially by Wyart and Cates~\cite{wyart2014discontinuous} 
in the context of shear thickening,
and later extended by others~\cite{guy2018constraint}.


These models address the activation and release,
under stress, of particle-level \emph{constraints}.
These are pairwise interactions that reduce the number of degrees of freedom 
along which particle motion can occur (e.g.\ through friction or adhesion), 
thus altering $\phim$.
A measure of the degree of constraint is encoded in a fraction of activated constraints $f$, 
which is a function of some component of the applied stress, say the shear stress $\sigma$
(some models instead use the particle pressure~\cite{wyart2014discontinuous}).
This function is usually given a sigmoidal shape around the typical stress $\sigma_\mathrm{c}$ 
needed to switch a microscopic constraint, such that $f(\sigma\ll \sigma_\mathrm{c}) \to 0$ 
and $f(\sigma\gg \sigma_\mathrm{c}) \to 1$.
In turn, $\phim$ depends on $f$.
In the absence of the microscopic constraint 
(for instance, absence of frictional contacts in the discussion in \secref{sec:repulsive}), 
$\phim(f=0) = \phim^{(0)}$, whereas when the constraints 
are fully mobilized $\phim(f=1) = \phim^{(1)}$ 
(for frictional contacts, $\phim^{(1)} < \phim^{(0)}$).
As sketched in~\figref{figure3}\,\figpanel{a},
$\phim$ controls the viscosity
through the divergence of $\suspvisc$ 
as, 
say, $\suspvisc \sim (\phim-\phi)^{-2}$ (but this specific form is not a requirement). 
$\suspvisc$ is then an implicit function of $\sigma$, 
interpolating between $\suspvisc = (\phim^{(0)}-\phi)^{-2}$ 
for $\sigma\ll \sigma_\mathrm{c}$ 
to $\suspvisc = (\phim^{(1)}-\phi)^{-2}$ 
for $\sigma\gg \sigma_\mathrm{c}$.
These models
have been quite successful at predicting steady state rheology~\cite{guy2015towards,royer_rheological_2016} 
and also inhomogeneous flows and instabilities~\cite{chacko2018dynamic,han2018shear}
(see also~\cite{nakanishiFluidDynamicsDilatant2012} for an independent model with similar structure).


This mechanism can be extended to 
arbitrary numbers of constraint types and activation (or release) stresses, 
leading to a wide range of possible predicted rheologies~\cite{guy2018constraint},
see examples 
in \figref{figure3}\,\figpanel{b}--\figpanel{e}.
Such Wyart--Cates like interpolation models can also be extended 
to tensorial constitutive laws~\cite{singh2018constitutive,baumgarten2019general}, 
and constant particle pressure rheology~\cite{wyart2014discontinuous,dong_analog_2017}.
While these models are motivated by a microscopic mechanism, 
they are in practice still phenomenological:
in addition to the phenomenological viscosity divergence law, 
the relation $f(\sigma)$ is usually postulated, although it is suggested 
that it is controlled by force chain physics~\cite{guy2015towards,royer_rheological_2016,ness2016shear}.
This opens the possibility of deriving $f(\sigma)$ from microscopic statistical models~\cite{mari_force_2019}.

\subsection{Two-phase description}

\label{sec:twophase}

Many flows involve inhomogeneous deformation, sedimentation or particle migration (\figref{figure2}\,\figpanel{i}), and one cannot describe the suspension as a single phase continuum. 
A two-phase description, coupling velocity, stress, and volume fraction fields for each phase, is necessary.
Such a model consists of conservation laws written separately for the solid and fluid phases, which one can obtain via local volumetric averages~\cite{jackson_locally_1997}. %
These involve a decomposition of the suspension stress into solid and fluid phase contributions (which are independently measurable~\cite{deboeufParticlePressureSheared2009,dbouk_shear-induced_2013}), 
as well as a coupling term in the form of an interphase drag.

These equations do not however form a closed system, as the stresses and interphase drag have no exact expression in terms of phase velocities or volume fraction.
Stresses are closed with a constitutive model, as discussed in the previous subsections.
A popular drag closure is the suspension balance model~\cite{nott1994pressure,nott2011suspension},
which consists of expressing the drag term as proportional to the relative velocity between the two phases.
This is in essence closely related to the approximation behind Darcy's law
for fluid flow through a porous solid phase.
These aspects and their implications are addressed in a recent review~\cite{guazzelli2018rheology}.

\subsection{Fluid mechanics}
\label{sec:inhomo}

Most flows of dense suspensions\,---\,especially in nature and in engineering\,---\,are not well-approximated by spatially homogeneous, rheometric conditions.
Instead, they typically involve spatial and temporal inhomogeneities in the stress, strain rate and volume fraction.
These situations are currently the focus of a significant research effort, 
drawing attention to several subtle aspects introduced below.
%


The bulk effective fluid description of dense suspensions presented 
in the previous sections is certainly a key tool to understand fluid mechanical phenomena, 
but it is not sufficient in itself.
Flows involving non-uniform conditions require a two-phase description as described above, 
and raise the question of the non-locality of constitutive models~\cite{gillissen2020modeling}.
Non-uniform systems can also be the result of instabilities, 
due to underlying non-monotonic flow curves, leading to banding~\cite{besselingShearBandingFlowConcentration2010}, 
or dynamic instabilities~\cite{Nagahiro_2016,rathee2017localized,chacko2018dynamic,nakanishiFluidDynamicsDilatant2012,saint-michelUncoveringInstabilitiesSpatiotemporal2018,ovarlez2020density,ratheeLocalizedTransientJamming2020}.
Shear-induced migration is also important in non-uniform conditions~\cite{morris2020shear,leightonShearinducedMigrationParticles1987,altobelliVelocityConcentrationMeasurements1991,Medhi_2019}.
Remarkably, there is a growing body of evidence that these instabilities are 
coupled to macroscopic deformation of the free surface, when present~\cite{hermes2016unsteady,ovarlez2020density,balmforth2005roll,darbois_texier_surface-wave_2020}.
%


The free surface poses a challenge as it represents a deformable confinement, constrained by both capillary forces and the incompressibility of the solvent.
This is in striking contrast to dry granular systems,
which are free to dilate.
The role of the free surface in the flow of dense suspensions is in general subtle, 
as the large particle size makes it easy for them 
to individually deform the solvent-air interface.
Curving a free surface on a particle radius scale $a$ requires a stress scaling as $1/a$ to counteract the effect of capillary forces, see \figref{figure2}\,\figpanel{j}. 
For micron-sized particles, this is easily achieved even in a rheometer, and particle-poking through the interface 
or a visible mattification of the surface are often reported~\cite{zhou2005theory}.
The free surface is then a nontrivial boundary condition for the fluid mechanical problem, 
in practice not well approximated by either constant volume 
nor constant pressure conditions~\cite{brownRoleDilationConfining2012}.
The crucial role of the grain scale deformation of the free surface is 
highlighted by situations where the interface slowly recedes, 
like unwetting or drying, which often leads to interface instabilities, particle deposition patterns and granulation~\cite{tang2000stability,eriksen2018pattern,ivesonNucleationGrowthBreakage2001,catesGranulationBistabilityNonBrownian2014}.
All these phenomena call for a careful modelling of the suspension free surface at the microscale, which is little-explored thus far,
at least in the context of dense suspension rheology.

\section{CLOSING REMARKS}
\label{sec:conclusion}

We have presented a perspective on the physics of dense suspensions.
Interpreting the macroscopic phenomenology of `real' suspensions and constructing continuum descriptions
on the basis of the simple microscopic arguments introduced here
will clearly be nontrivial in most cases.
Nonetheless, what we have presented are foundational concepts from which descriptions of more complex systems can be assembled.
Doing so will require further study in many areas, some of which we propose below as Future Issues.
Given the ubiquity of dense suspensions both as a model system in condensed matter physics and as an engineering material,
there is no doubt that they will remain a topic of research and debate in future.

\begin{issues}[FUTURE ISSUES]

\begin{enumerate}
\item Can suspensions of non-idealized particles (aggregates, fibres, rods, with arbitrary polydispersity) be approached by analogy to nearly monodisperse spheres, that is, only adjusting the values of key parameters like $\phim$ and the exponent $\beta$?

\item How far can the insights presented here be carried to the physics of suspensions with non-Newtonian solvents? Recent work suggests that in some cases the rheology can be captured by an effective local shear rate in the solvent, controlled by geometry close to jamming~\cite{dagois-bohyRheologyDenseSuspensions2015}.

\item The sensitivity of the jamming transition to the nature of particle contacts calls for more attention to, in particular, the physics of contacts with a few surface asperities (as opposed to the multi-contact interface paradigm of friction between macroscopic particles)~\cite{lobry2019shear}. 

\item The statistical physics connecting microscopic details to flow close to jamming is limited to frictionless spheres~\cite{during_length_2014,degiuli_unified_2015}. Is the floppy mode paradigm relevant with friction present~\cite{trulsson_effect_2017}?

\item Suspensions with non-rigid constraints often have elastic structures. Can viscous continuum models such as those described above be augmented to account for these?

\item The boundary condition provided by free surfaces plays a major role but is poorly understood because it often involves particle-size curvature. What aspects of interfacial physics (such as contact angle and roughness anchoring) matter most at the boundary?

\item More generally, to what extent do the simplifications made here (limited role of hydrodynamics, rheometric conditions, etc.) need to be relaxed in order to describe real flows which may involve instabilities and secondary structures?

\end{enumerate}
\end{issues}

\section*{DISCLOSURE STATEMENT}
The authors are not aware of any affiliations, memberships, funding, or financial holdings that
might be perceived as affecting the objectivity of this review. 

\section*{ACKNOWLEDGMENTS}
The authors thank M. Cates and E. Guazzelli for commenting on the manuscript.
C.N. acknowledges support from the Royal Academy of Engineering under the Research Fellowship scheme.
R.S. acknowledges the support from the Wenzhou Institute, University of Chinese Academy of Sciences,
under Grant No.\,WIUCASQD2020002.

\bibliographystyle{ar-style4}
\bibliography{library}

\end{document}